# Affordable Virtual Reality System Architecture for Representation of Implicit Object Properties

Stoyan Maleshkov[1], Dimo Chotrov[1]

[1] Virtual Reality Lab, Technical University Sofia, Sofia, 1000, Bulgaria

**Abstract**
A flexible, scalable and affordable virtual reality software system architecture is proposed. This solution can be easily implemented on different hardware configurations: on a single computer or on a computer cluster. The architecture is aimed to be integrated in the workflow for solving engineering tasks and oriented towards presenting implicit object properties through multiple sensorial channels (visual, audio and haptic). Implicit properties represent hidden object features (i.e. magnetization, radiation, humidity, toxicity, etc.) which cannot be perceived by the observer through his/her senses but require specialized equipment in order to expand the observer's sensory ability. Our approach extends the underlying general scene graph structure incorporating additional effects nodes for implicit properties representation.

## 1. Introduction

Virtual Reality (VR) has already passed the initial phases of inflated expectations and disillusionment typical for every new technology, moving nowadays to a period of clarification and integration in everyday human activities. In the field of engineering applications the ability to explore the design of a product in a computer generated immersive environment and to verify its functionality using a virtual prototype rather than building up a physical one gives considerable advantage over the competition reducing time-to-market, especially for small and medium enterprises. Research results have demonstrated that for common data exploration and analysis tasks the immersive systems have provided notable insight and considerable time savings, becoming a part of the research workflow. Recent development in the IT sector have significantly reduced the cost of all major physical parts of a virtual reality system: stereoscopic displays and projectors, audio and haptic devices, input and rendering hardware can fit in a reasonable prized project budget. However, the software component of a virtual reality system still remains quite expensive, especially if delivered as commercial product. The effort to create open source software solution to support various types of applications has led to the development of several VR integration libraries and VR systems but common support and consistency are still missing or do not comply with the user requirements. This paper describes our research to design and implement low cost VR system architecture to be integrated in the workflow of solving engineering tasks and oriented towards presenting implicit object properties through multiple sensorial channels (visual, audio and haptic). Basic activity when exploring objects in VR environment is to examine and verify the functionality of the design by evaluating object properties. Special attention in our research is given to the implicit properties which represent hidden object features (i.e. magnetization, radiation, humidity, toxicity, etc.) which cannot be perceived by the observer through his/her senses but require specialized equipment in order to expand the sensory range of the observer's different sensory channels. Until now, especially in virtual engineering, representation of implicit features is performed mainly through their substitution with primitives perceived through the human visual channel, i.e. temperature or pressure distribution is presented through color codes or icons. Enhancing the representation of implicit properties with other senses besides seeing for the human-computer interaction and specifically for information presentation will significantly increase productivity in the process of solving engineering problems. Combining explicit and implicit object properties and mapping to multiple stimuli to make the user simultaneously aware of different features of a virtual object will increase the ability of the user to perceive and evaluate several object features at the same time.

## 2. Related work

Several open software system architectures for VR integration have been developed during the last decade as an affordable alternative to the commercial VR software packages. One group of software solutions are designed for building large high-end VR applications e.g. VRJuggler [2] or AVANGO [6], which describe a single common, modular architecture for different devices. The generic VR software system architecture VR$^2$S [11] defines a high-end abstraction for building virtual environments. In contrast to most other graphics systems, the rendering is based on independent API providing the opportunity to switch between different rendering systems at runtime without the need for redesign of the application source code.

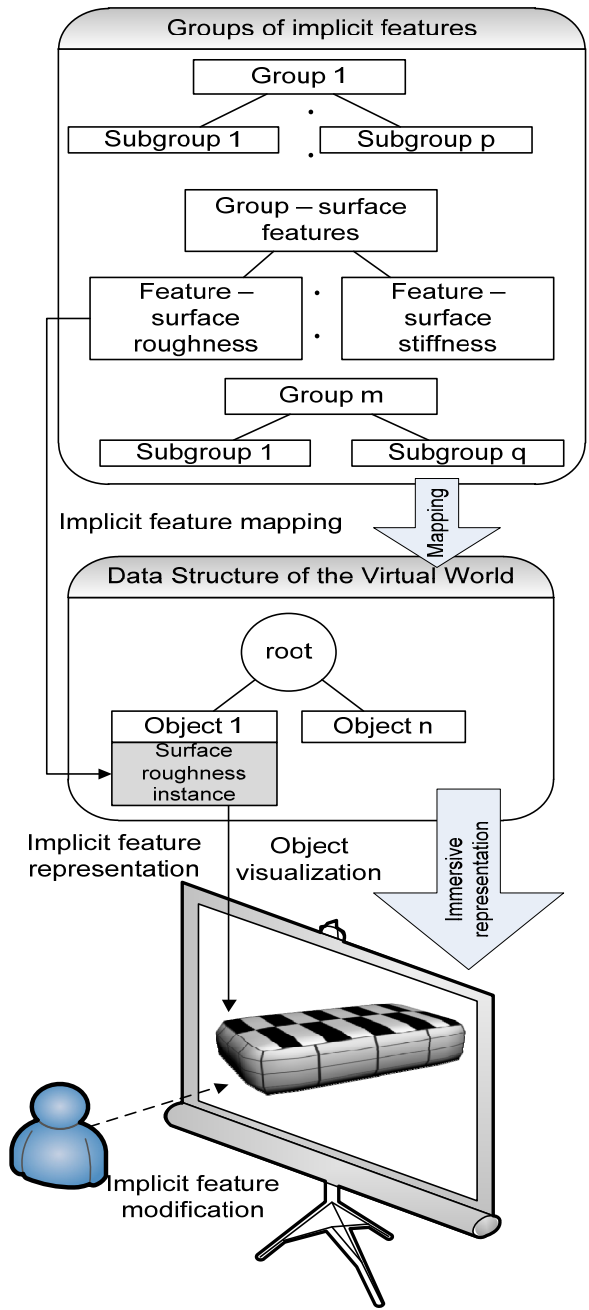

Fig. 1. Enhanced representation of object properties in virtual environment implementing implicit features – represent the implicit feature surface roughness through the visual channel applying different textures on the corresponding surfaces

The development of low cost VR system architecture is discussed in [4] and [7]. Affordable portable solutions are presented in [7] and [10]. To our opinion the lack of relevant software support and consistency limits the penetration of the VR software architectures described above. We have created our solution based on the NVidia SceniX [9] receiving reasonable product support and update policy.

The common method for representation of the virtual scene is to use a scene graph, first proposed in 1992 by Strauss and Carey in their 3D Graphics Toolkit [13]. The scene graph is a graph-type data structure which simplifies and improves the performance of the visualization as well as interaction and transformation of the objects in the virtual scene. The scene graph contains nodes describing geometrical information for the objects in the scene, materials that should be used when rendering the objects, transformations for spatially arranging the objects, as well as nodes for description of cameras and lighting. The scene graph can be used to avoid data redundancy by referencing the same geometry from two or more different group nodes.

Using the scene graph as a basis, the underlying data structure can be extended by the design of the VR system architecture to incorporate additional data. This approach can be applied for adding audio [3] and [12] in the VR system as well as haptic signals allowing for multimodal human computer interaction and increasing the sense of immersion in the computer generated environment.

The concept of implicit object features and their representation in VR environment has been considered and demonstrated in [1]. The basic approach uses two separate data structures: one for representation of objects in the virtual world and another one for description of the implicit features, and implements mapping and multimodal representation during the rendering as shown on Fig. 1.

## 3. Integration of Implicit Features in a Virtual Environment

For the representation of implicit features in a virtual environment different stimuli called *Effects* are used. Our research focuses on three types of effects: audio, visual and haptic, divided by the users' senses they stimulate. The effects can be combined to represent different implicit features of the same object e.g. using audio effect to denote the presence of a magnetic field around (and in) an object and haptic for describing the temperature distribution along the surface of the same object.

For the integration of implicit feature representation in a virtual environment a proper mechanism has to be adopted for assigning implicit features to a virtual object, for storing information about these features and for describing the representation of these features in a multimodal user interface environment.

3.1 Scene Graph Extension

Using the general structure of a scene graph as a base we extend it so that the new scene graph includes also data needed for the multimodal representation of implicit features. This allows the presentation of implicit features to be easily handled during a scene graph rendering process. User interaction requiring implicit feature presentation can also be detected by a scene graph traversal process. Two additional node types have been designed and included in the scene graph structure to store the necessary data. The first type describes nodes which can exist independently – e.g. the *AudioNode* as shown on Fig. 2.

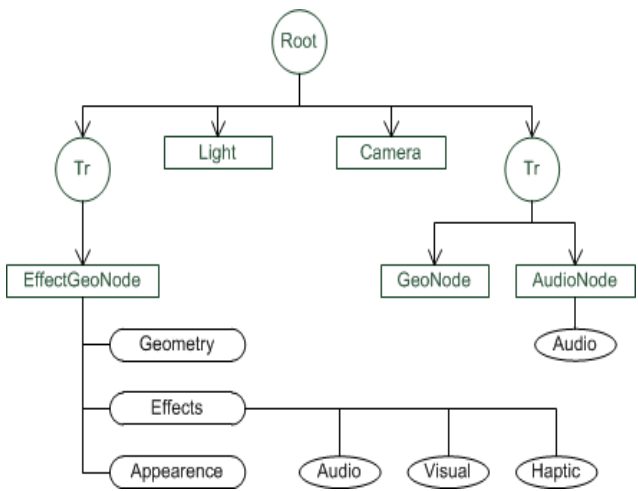

Fig. 2. Scene graph containing the traditional nodes extended with the new node types: AudioNode and EffectGeoNode. Tr denotes TransformNode and GeoNode is a node containing a virtual object.

These nodes can be bound directly to a transformation positioning a sound source in the virtual world. And the second type – nodes that hold geometric data and also contain information about different effects that can be used to react to user interaction with an object or part of an object – e.g. the *EffectGeoNode* on Fig. 2.

3.2 Implicit Feature Mapping

A specialized configurator is provided to access the transform and geometry nodes contained in the scene graph of a virtual scene allowing the mapping of implicit features to virtual objects. With this tool the user can select individual nodes from the scene graph and then assign and adjust different effects to be used to represent the implicit features of the node (Fig. 3).

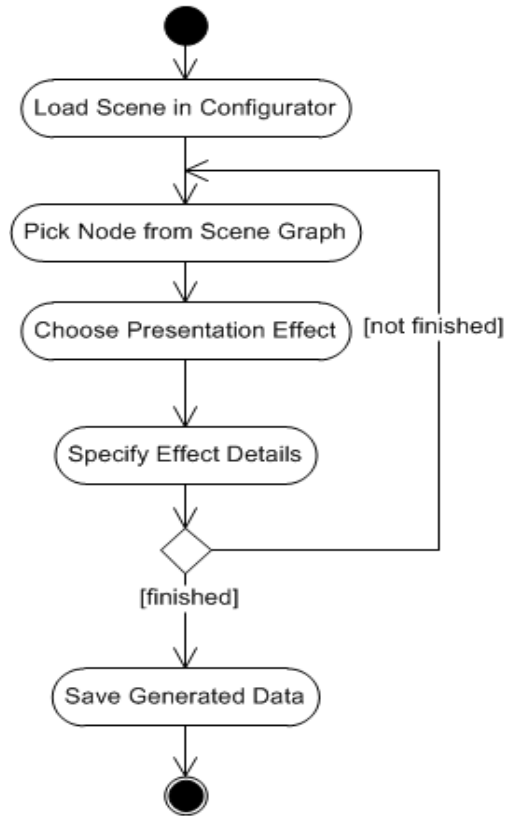

Fig. 3. Mapping implicit feature presentation effects to objects in a virtual scene.

During the process of mapping the initial graph describing the scene is modified to reflect the changes made by the user. Referring back to Fig. 2 two different actions can be performed: either a child *AudioNode* is added to a *Transform* node or a *GeoNode* is replaced by an *EffectGeoNode*. In the latter case the geometric data of the *GeoNode* is copied to the *EffectGeoNode* so that the new node contains the geometry describing the virtual object as well as the data specifying the selected effect. The result is a new scene graph used for the representation of the virtual scene incorporating the implicit features of the objects.

## 4. VR Software Architecture

The general structure of the proposed affordable virtual reality system architecture follows a distributed software model with one management node, called producer (see

Fig. 4) and one or more consumer nodes. The producer provides a graphical user interface and also allows the integration of additional input modules, for example modules for user input through a tracking system and gesture recognition. In general the producer generates information based on user input that is to be presented to the user by the consumers.

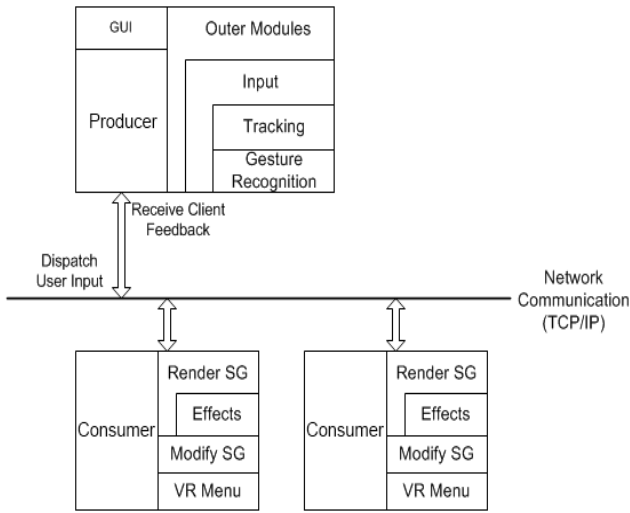

Fig. 4. Proposed software architecture of the virtual reality system.

The consumers execute the actual traversal and rendering of the scene graph, including the presentation of effects describing the implicit features mapped to virtual objects. They are also responsible for keeping the scene graph up to date according to changes caused by user interaction. The consumers expose a VR menu which can be turned on by the user in order to perform different actions or to specify how specific user actions are to be interpreted by the system e.g. switching from exploration to selection or editing mode.

The communication between the producer and the consumers is organized over TCP/IP. The producer maintains connection with each of the consumers and uses it to send packets informing the consumer of specific user commands (i.e. movement, gestures) and to receive consumer feedback.

4.1 Modules of the VR Software System

The system is decomposed into a number of modules responsible for specific tasks, organized in packages as shown on Fig. 5.
The *Producer* application is responsible for getting input from the user. It provides a graphical user interface which can be used, for example, to load a scene, to modify the current view point or to play an animation. The *Producer* accesses the *DTRackUDPReceiver* module to receive information about user movements from a tracking system.

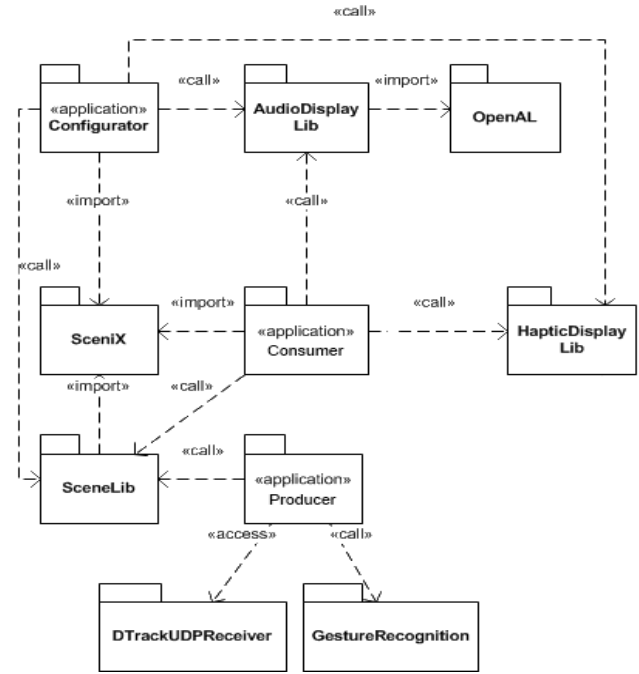

Fig. 5. Decomposition of the proposed VR system architecture into packages.

It can then use this information to check if the user has performed a gesture by calling the *GestureRecognition* module. When the *Producer* receives a command from the user it sends an appropriate packet to all *Consumer* applications connected to it.

The *Consumer* application renders the virtual scene from a specified view point. It uses the *SceniX* API for scene visualization and can use the *AudioDisplayLib* and *HapticDisplayLib* for the audio and/or haptic presentation of implicit features.

The *Configurator* application is used for mapping of implicit features to objects from a virtual scene. It relies on the *SceneLib* library to load the base scene graph (without implicit features) which is displayed to the user during the mapping process. Afterwards the *Configurator* generates an implicit feature mapping description file which is used by *SceneLib* to create the final scene graph for representation of the virtual scene incorporating objects' implicit features. The *Configurator* uses the *SceniX*, *AudioDisplayLib* and *HapticDisplayLib* libraries to show a preview of the virtual object presented with its implicit features.
The audio and haptic display libraries are contained in separate packages – *AudioDisplayLib* and

*HapticDisplayLib*. This allows the output modules to be attached independently to any *Consumer* application. For example in a multi-consumer configuration one consumer can be responsible for audio output, another for haptic output, while in the same time both consumers are responsible for video output. Or haptic output can be provided by a third consumer, which does not render visual information but is dedicated to haptics only.

The *SceneLib* module is responsible for tasks like loading a virtual scene from a file into a scene graph and performing actions on the scene graph (or part of it) - for example determining ray intersections, object picking. *SceneLib* uses the *SceniX* API for generating the base scene graph for a virtual scene. It can then use the input from the *Configurator* application to modify the scene graph so, that it contains also the nodes needed for the implicit feature presentation. *SceneLib* is used by all the applications in the system to receive a scene graph for a given virtual scene. It is important here to mention that when an application calls *SceneLib* to create the scene graph for it, it also sends to *SceneLib* information for the presentation of which effects is it responsible. As a consequence when creating the scene graph *SceneLib* can skip effects for which the application is not responsible. The result is a scene graph containing only nodes of interest for the application – i.e. if an application is not used to present haptic effects its scene graph will not contain such effects.

The proposed VR software architecture uses 2 external packages: *SceniX* [9] is a scene graph software development kit designed by NVIDIA which aims at representing virtual scenes interactively with high quality by taking advantage of the latest improvements in graphics hardware. Among its features it provides real time ray tracing and the possibility for easy integration of a physics library. *SceniX* is becoming more popular and, as it is supported by one of the main graphics hardware and software manufactures, has a good chance to become a leading and widely accepted scene management engine. *OpenAL* [8] is a cross-platform 3D audio API. It can simulate mobile or static audio sources in a virtual space that can be heard by a listener whose position is defined in the same space. *OpenAL* is supported by Creative Labs.

4.2 Representation of Implicit Features

The diagram on Fig. 6 describes the classes used for the representation of the new scene graph nodes and for the presentation of the implicit feature effects stored in them. To keep the diagram simple and readable we have concentrated mainly on the presentation of implicit features through the audio channel as the classes and handling of the other two types of effects are similar.

The classes starting with *nvsg::* are part of the SceniX [8] API, which our scene description and management is based on. The class *nvsg::Node* is the basic type for a node in the scene graph. It is used as a base class for the other node classes. The *AudioNode* class is inherited from it describing a simple node, which only stores additional audio data to be played when rendering the scene graph. The *EffectGeoNode* class is inherited from n*vsg::GeoNode* which represents a node containing geometrical information of an object. This information can be used to determine the exact surface element of a virtual object pointed to by the user, which can affect the way an implicit feature is presented to the user – e.g. to distinguish between surface areas of a virtual object with different temperatures.

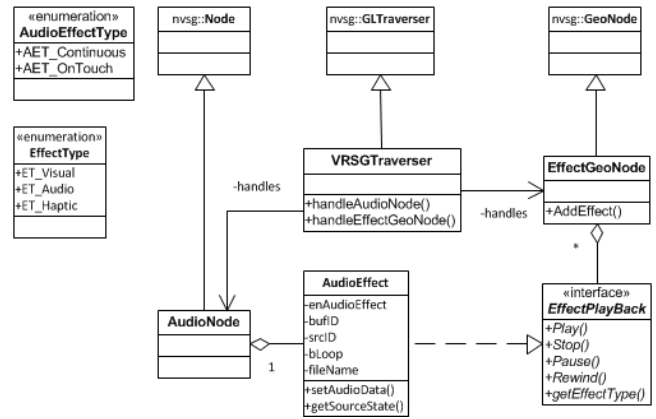

Fig. 6. Classes designed to represent implicit features related to audio data.

In order for the system to be able to interpret the additional information in the newly defined scene graph node types actions have to be defined that can handle these nodes. For this purpose an interface called *EffectPlayback* has been designed. The interface provides abstraction allowing the action handling the node to execute the effect without knowing the exact type of the effect. This allows the *EffectGeoNode* class to contain a list of effects of type *EffectPlayBack*, instead of storing separate lists for every effect type. The *getEffectType* method of the interface helps to identify the actual type of the stored effect, as described by the *EffectType* enumeration.

The *VRSGTraverser* class performs the rendering traversal of a scene graph, including the presentation of implicit features contained in the new node types. It is inherited from the SceniX *nvsg::GLTraverser* class, which handles the rendering of the object geometry, so that only the new actions for handling the presentation of implicit features have to be defined – *handleAudioNode* and

*handleEffectGeoNode*. The two methods only need to make a call to the *Play* method of the *EffectPlayBack* interface, implemented by the respective effect, and the latter handles the implicit feature presentation.

The *AudioEffect* class is used for storing the actual data specifying the mapped implicit feature effect. The *bufID* and *srcID* attributes store OpenAL [7] source and buffer identifiers needed for sound playback. The *enAudioEffect* attribute stores an *AudioEffectType* enumeration value describing the actual type of the audio effect. *AET_Continuous* denotes an audio effect that is being continuously played – such effects are handled during the scene graph rendering traversal. *AET_OnTouch* are audio effects that are played on user interaction with a virtual object. The *setAudioData* method initializes the *bufID* and *srcID* attributes needed for the presentation of the audio effect. There are also two overloaded methods: one allowing the audio data initialization and memory allocation from a file stream and another one using already allocated memory for specifying dynamically generated audio data.

## 5. Implementation

To explore a scene with implicit features in a virtual reality environment the user first needs to create the mapping of effects describing how the implicit features of the objects in the scene are going to be presented. For the purpose the user has to start the Configurator application and open the scene in it. The Configurator provides a Tree-View control showing a hierarchical view of the base scene graph and a list of effects that can be assigned to the nodes of the scene graph. The user selects a node from the scene graph and chooses one or more effects to be assigned to the object. When the mapping is accomplished the user has to save the resulting description - a file containing the object-to-implicit feature mapping is generated. Then the user starts the Producer application which on its turn starts the Consumer application(s). The configuration of the virtual reality system is determined by a configuration file describing number of consumers, addresses of the producer and consumer machines to be used, common storage location (from which the scene is to be loaded), details about tracking devices, etc. The configuration file is also used to determine each application and machine for what kind of output is responsible. After the system is started the user specifies the file containing the virtual scene and the file generated from the Configurator application with the implicit feature mapping. The system loads the specified virtual scene together with its implicit feature representation and the user can start exploring it in the generated virtual environment.

Using the distributed software model a scalable system can be created that can make use of different number of consumers allowing the implementation of the system in different hardware configurations. Different number of consumers can connect to the producer depending on the hardware configuration in use. For example, in a single consumer configuration (Fig. 7a) the producer and the consumer can run on one and the same computer. In this case the stereo visualization is performed with proper hardware: quad buffered video card and a 3D capable display. In single consumer configuration the user can also interact directly with the consumer application using mouse and keyboard to perform tasks like picking, object translation, etc. A two consumer configuration can be used for power-wall visualization (Fig. 7b). In such a case the system could run on a cluster of three computers – one for the producer and two for the consumers. Each consumer is responsible for generating the image for one of the eyes and both images are projected on the power-wall in order to produce passive stereoscopic visualization.

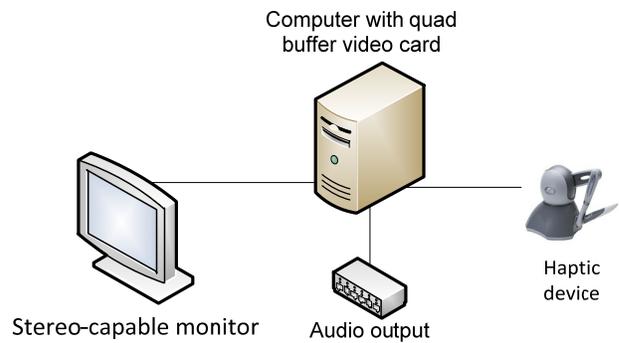

Fig. 7a. Sample configuration for virtual reality presentation in single consumer configuration

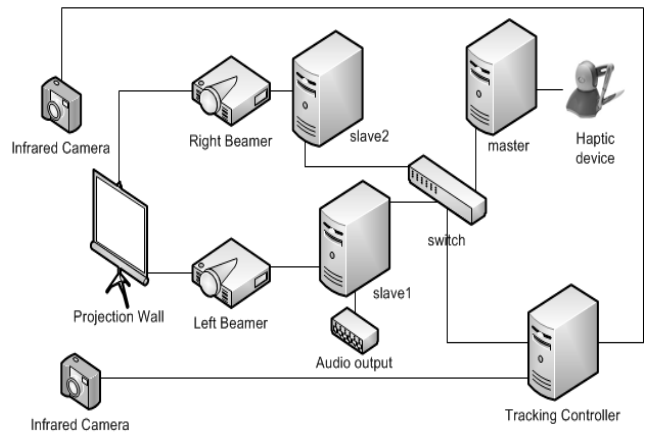

Fig. 7b. Sample configuration for virtual reality presentation on a power-wall.

Fig. 7b shows a sample virtual reality configuration applying one screen rear passive stereoscopic projection as the one used at our Virtual Reality Lab. The configuration consists of a cluster of three computers - one master and two slaves, infrared optical tracking system, audio system and a haptic device. In this case the producer application runs on the master computer, two consumers run each on every slave and one more consumer application runs on the master computer. The producer receives tracking information and user input. The two consumers on the slave computers render the images of the virtual world for the left and right eyes needed for the stereo visualization. In this case a haptic device is connected to the master computer and will be managed by the consumer application on that computer, whereas the consumer on one of the slave computers is responsible for the audio output. The cost of such configuration is about 30 thousand euro (Christie DLP projectors and ART tracking). Single computer desktop stereo visualization solution with NVidia 3D Vision can be implemented at a cost of about 2 thousand euro.

## 6. Conclusion

In this paper we propose an affordable, scalable and flexible architecture for virtual reality representation of implicit features and a procedure for mapping implicit features to objects from a virtual scene. The proposed architecture can run on different hardware configurations – on a single computer or on a computer cluster. It also allows tuning of the used hardware system by providing the ability to attach audio and haptic displays to different machines which allows for balancing the performance of the system, especially in cases when different machines are used to build up a cluster.

The scalability of the system comes as a consequence of the applied distributed software model. It is relatively easy to add more consumers responsible for other views of the scene (i.e. for visualization in a multiple screen VR system).
The modular architecture of the system allows for easy improvement, replacement or addition of modules, for example to add support for another haptic device or create another effect for implicit feature representation. This makes the system flexible and easy to maintain.

**Acknowledgments**

The authors wish to thank for the support of the National Science Found at the Bulgarian Ministry of Education, Youth and Science received through grant DDBY02/67-2010.

**References**
[1] A. Bachvarov, S. Maleshkov, D. Chotrov, J. Katicic, "Immersive Representation of Objects in Virtual Reality Environment Implementing Implicit Properties", Springer (2011), 4-th International Conference on Developments in eSystems Engineering - DeSE 2011, pp. 587-592.
[2] A. Bierbaum, C. Just, P. Hartling, K. Meinert, and A. Baker, "VR Juggler: A Virtual Platform for Virtual Reality Application Development", VR-01, IEEE, 2001, pp. 89-96
[3] K-U. Doerr, H. Rademacher, S. Huesgen, and W. Kubbat, "Evaluation of a Low-Cost 3D Sound System for Immersive Virtual Reality Training Systems", in IEEE Transactions on Visualization and Computer Graphics, 2007, vol. 13, pp. 204-212
[4] T. Fellmann, and M. Kavakli, "VaiR: System Architecture of a Generic Virtual Reality Engine", in Proceedings of the International Conference CIMCA '05,2005,vol.2, pp.501-506
[5] M. Kalkush, and D. Schmalstieg, "Extending The Scene Graph With A Dataflow Visualization System", in VRST'06 Proceedings of the ACM symposium on Virtual reality software and technology, 2006, pp. 252-260
[6] R. Kuck, J. Wind, K. Riege, M. Bogen. Improving the AVANGO VR/AR Framework |- Lessons Learned. Virtuelle und Erweiterte Realität: 5. Workshop der GI-Fachgruppe VR/AR, Magdeburg, 25.-26.09.2008, pp. 209-220.
[7] M. Külberg, J. C. de Oliveira, and P.F.F. Rosa, "MiniVR: Low Cost VR Projection System", in XIII Symposium on Virtual Reality, 2011, pp. 134-143
[8] OpenAL, http://connect.creativelabs.com/openal/default.aspx, Last accesses 02.03.2012
[9] SceniX, http://developer.nvidia.com/scenix-details, Last accesses 02.03.2012
10] W. Sherman, P. O'Leary, E. Whiting, S. Grover, and E. Wernert, "IQ-Station: A Low Cost Portable Immersive Environment", in ISVC 2010, Part II, 2010, LNCS 6454, pp. 361-372
[11] F. Steinicke, T. Ropinski, and K. Hinrichs, "A Generic Virtual Reality Software System's Architecture and Application", in Proceedings of ICAT '05, 2005, pp. 220-227
[12] T. Stockman, L. V. Nickerson, and G. Hind, "Auditory graphs: A summary of current experience and towards a research agenda", 11-th Int. Conf. on Auditory Display, 2005
[13] P. Strauss, and R. Carey, "An Object Oriented 3D Graphics Toolkit", in Computer Graphics Proceedings of SIGGRAPH 92, 1992, Vol. 26, pp. 341–349

**Stoyan Maleshkov** is associate professor of computer aided engineering and computer graphics (1990) at the Technical University (TU) of Sofia, Bulgaria. He has Eng. degree in system and control engineering (1975), master in applied mathematics (1977) and PhD in computer aided system design (1981), all received from the TU of Sofia. Fulbright scholar (1989–1990) at Interactive Modeling Research Lab, Louisiana State University, Baton Rouge, USA. Department chair (2000-2004) and vice dean (2004-2008), both at the TU Sofia. Since 2008: Head of the Virtual reality lab, TU of Sofia. Associate professor of computer graphics at the New Bulgarian University, Sofia - as a second job (2000).

**Dimo Chotrov** has received BSc. (2007) and MSc. (2009) degrees in computer systems and technologies from the Technical University of Sofia. Currently he is PhD student, acting at the Virtual reality lab. Member of the IEEE.